\documentclass[letterpaper,english,reprint, aps]{revtex4-1}
\usepackage[T1]{fontenc}
\usepackage[utf8]{inputenc}
\usepackage{color}
\usepackage{amsmath}
\usepackage{amssymb}
\usepackage{graphicx}

\makeatletter

\usepackage[colorlinks = true,
            linkcolor =red,
            urlcolor  =magenta,
            citecolor = blue,
            anchorcolor = blue]{hyperref}

\makeatother

\usepackage{babel}
\begin{document}
\title{Single-photon-added coherent state based postselected weak measurement}
\author{Yi-Fang Ren }
\author{Yusuf Turek}
\email{yusuftu1984@hotmail.com}

\affiliation{$^{1}$School of Physics, Liaoning University, Shenyang, Liaoning
110036, China}
\date{\today}
\begin{abstract}
We investigated precision measurements in a two-level system coupled
to a single-photon-added coherent state (SPACS) under postselection
measurement. We analyzed strategies for improving measurement precision,
including parameter estimation and the signal-to-noise ratio (SNR)
in postselected weak measurements using the photon statistics of SPACS
as the meter. Our results demonstrate that SPACS-based postselected
weak measurements can outperform conventional measurement schemes
in terms of precision. Additionally, we explicitly introduced an alternative
weak measurement method commonly applied in dispersive light-atom
interactions. Our work offers a new way for addressing fundamental
issues in quantum precision measurement based on photon statistics,
and it provides a method for extracting the phase and phase shifts
of radiation fields through the weak values of system observables.
\end{abstract}
\maketitle

\section{Introduction}

Quantum measurement is a fundamental aspect of quantum mechanics and
is essential to understand the microscopic world. Von Neumann’s projective
strong measurement and weak measurement proposed by Aharonov et al.
\citep{PhysRevLett.60.1351} are two central approaches within the
study of quantum measurement problems. Both measurement models can
be applied in quantum metrology, enabling the extraction of desired
system information using appropriate methodologies. However, in recent
years, the weak measurement theory proposed by Aharonov et al. has
attracted considerable attention due to their unique features in a
wide variety of practical applications in precision measurements compared
to strong projective measurements \citep{PhysRevLett.117.230801,02487,PhysRevLett.125.080501,PhysRevLett.134.080802}.
In the context of weak measurements, postselection allows for the
emergence of the weak value, which can fall outside the range of the
observable eigenvalues of the system \citep{Aharonov1964TIMESI,Aharonov2008,RN19}. 

The weak value can become arbitrarily large by appropriately selecting
the initial and postselected system states. This phenomenon, known
as weak value amplification (WVA), has emerged as a powerful technique
in quantum metrology for amplifying small physical effects across
various research areas, the detection of the spin Hall effect \citep{doi:10.1126/science.1152697,PhysRevA.85.043809},
phase shifts \citep{PhysRevLett.105.010405,PhysRevLett.111.033604,76312}
and nonlinearities \citep{PhysRevLett.107.133603,Wang_2019}. For
detailed discussions on applications of weak measurements in various
research fields, the reader is referred to \citep{KOFMAN201243,Tamir2013,RevModPhys.86.307}
and the references therein.

WVA has several practical applications in quantum metrology, as it
enhances measurement precision through postselection, as investigated
in Ref. \citep{11,XU2024100518}. However, achieving large amplification
through postselection typically results in a lower success probability,
which can reduce its metrological advantage. Although postselected
weak measurements employing WVA have successfully addressed several
precision measurement challenges \citep{PhysRevLett.102.173601,62,PhysRevA.82.063822},
controversy remains regarding the optimal WVA strategy for maximizing
measurement precision \citep{PhysRevLett.115.120401,PhysRevA.106.022619,PhysRevA.88.042116,PhysRevLett.114.210801,GUO2021104868}.
A key issue in quantum metrology is how to enhance parameter estimation
precision efficiently, using practical and cost-effective resources.

The widely used benchmarks to quantify the accuracy of the unknown
estimated parameters are signal-to-noise ratio (SNR) and quantum Fisher
information (QFI) \citep{PhysRevA.87.012115,PhysRevA.88.042116,PhysRevLett.112.040406,PhysRevLett.115.120401,PhysRevLett.114.210801,PhysRevX.4.011031,PhysRevA.102.042601,PhysRevA.106.022619}.
In Ref. \citep{PhysRevA.88.042116,PhysRevLett.114.210801}, the results
suggest the negative conclusion that weak measurement cannot improve
parameter estimation based on the Fisher information (FI) conditioned
on successful postselection. However, in recent work \citep{PhysRevA.106.022619},
WVA metrology was investigated using an optical coherent state as
a meter, demonstrating that the FI in the WVA-based scheme can surpass
that of conventional measurements not employing postselection. This
interesting result contradicts the conclusion of Ref. \citep{PhysRevLett.105.010403},
which claimed that for a linear detection scheme without postselection,
achieving metrological resolution beyond the coherent state limit
is genuinely due to a nonclassical effect. In other studies \citep{PhysRevLett.115.120401,PhysRevA.92.022109,GUO2021104868,RN17},
improvements in the precision of WVA measurements, characterized by
a higher SNR compared to conventional measurements, were confirmed
using nonclassical meter states.

A coherent state is a convenient choice due to its semi-classical
nature. However, in Ref. \citep{PhysRevA.106.022619} demonstrated
that higher WVA-QFI compared to conventional measurements is achieved
only in the regime of stronger measurement strength. This result raises
the question of whether it is possible to achieve a similar advantage
using a coherent state in regimes of weaker interaction strength.
In the context of weak measurements, we may consider measurement interaction
Hamiltonians of the von Neumann type $g\hat{A}\otimes\hat{P}$ or
$g\hat{A}\otimes\hat{X}$ where $\hat{A}$ represents the observable
of the measured system, and $\hat{P}$ and $\hat{X}$ denote the canonical
momentum and position operators of the meter, respectively. These
operators satisfy the canonical commutation relation $[\hat{X},\hat{P}]=i$. 

In recent studies \citep{RevModPhys.82.1155,PhysRevLett.114.210801,PhysRevA.99.022114,PhysRevA.106.022619},
quantum measurement based on photon-number interactions of the type
$g\hat{A}\otimes\hat{n}$ has been considered. Here, $\hat{n}=\hat{a}^{\dagger}\hat{a}$
is denotes the photon-number operator and its canonically conjugate
pair called ``phase operator''. Even though there have no well-defined
phase operator, Paul Dirac firstly postulated a phenomenological uncertainty
relation for the electromagnetic field phase and photon number, $\Delta n\Delta\varphi\geq\frac{1}{2}$.
Hence, similar to the position and momentum operators, the photon-number
operator $\hat{n}=\hat{a}^{\dagger}\hat{a}$ and phase operator $\hat{\varphi}$
obey the the canonical commutation relation $[\hat{n},\hat{\varphi}]=i$
\citep{PhysRevA.39.1665,Barnett01011989,RN15}. Nevertheless, it is
natural to consider whether a weak measurement proposal for the interaction
of the type $g\hat{A}\otimes\hat{n}$, analogous to the original concept
introduced by Aharonov et al., can be formulated.

To address the first question, one can consider nonclassical meter
states, as they can potentially enhance the precision of parameter
estimation. Our answer to the second question is affirmative.

To investigate the above two problems, we choose the single-photon-added
coherent state (SPACS) as a meter. The SPACS is defined as $(1+\vert\alpha\vert^{2})^{-\frac{1}{2}}\hat{a}^{\dagger}\vert\alpha\rangle$,
where $\hat{a}^{\dagger}$ is the photon creation operator and $\vert\alpha\rangle$
denotes a coherent state. The motivation for choosing SPACS as a meter
is twofold: (i) although SPACS differs from the coherent state by
only a single photon, its photon statistics differ significantly,
exhibiting nonclassical behavior for small $\vert\alpha\vert$; and
(ii) results obtained with SPACS smoothly reduce to the coherent-state
results in the limit $\vert\alpha\vert\gg1$. 

In this paper, we present a theoretical analysis of postselection
measurement using the SPACS as a meter. We focus on comparing postselected
weak measurements with conventional measurement strategies and discuss
the advantages of the WVA strategy in terms of state distance and
SNR. Additionally, we introduce a photon-statistics-based postselection
measurement scheme and explore photon statistics to characterize its
performance. Finally, we numerically investigate the uncertainty relations
for the photon-number and phase operators and the metrological advantage
provided by photon statistics, particularly related to weak value.

The remainder of this paper is organized as follows. In Sec. \ref{sec:2},
we investigate the FI associated with postselection measurements and
discuss its implications for parameter estimation. This section consists
of three subsections. The first subsection presents the QFI obtained
using the conventional measurement strategy and the WVA strategy.
The second and third subsections investigate specific implementations
of the WVA strategy, focusing respectively on photon-number measurement
and field-quadrature measurement. In Sec. \ref{sec:4}, we perform
analyses of state fidelity and SNR to demonstrate the benefits of
postselection for enhancing measurement precision and accuracy under
weak measurement conditions. In Sec. \ref{sec:5}, we explore an alternative
weak measurement strategy based on photon statistics, focusing on
how the weak values of the measured system's observable influence
average photon number and phase distributions. We also discuss the
uncertainty relation between photon-number and phase operators. Finally,
in Sec. \ref{sec:6}, we summarize our conclusions.

\section{\label{sec:2}Quantum Fisher Information with different methods}

In previous work Ref. \citep{PhysRevA.106.022619}, quantum measurements
based on photon-number interactions have been explored, particularly
in the context of WVA, where a coherent state was used as a meter
to enhance parameter estimation precision. However, in our work, we
introduce a SPACS as the meter, which offers significant nonclassical
features that are expected to enhance the precision of weak measurements
in comparison to coherent-state-based measurements. In sections \ref{sec:2},
we adopt a structure and notation are largely based on \citep{PhysRevA.106.022619},
ensuring a clear comparison between the two approaches. This choice
allows us to explore the impact of nonclassical meter states on measurement
precision, particularly in the weak measurement regime, where SPACS-based
measurements are expected to offer advantages over coherent-state-based
measurements.

We assume a two-level quantum system (qubit) with states $\vert g\rangle$
and $\vert e\rangle$, coupled to a SPACS. The interaction Hamiltonian
can be written as:
\begin{equation}
\hat{H}=g\hat{\sigma_{z}}\otimes\hat{n},\label{eq:1}
\end{equation}
where $\hat{\sigma}_{z}=\vert e\rangle\langle e\vert-\vert g\rangle\langle g\vert$
is the Pauli operator, $\hat{n}=\hat{a}^{\dagger}\hat{a}$ denotes
the photon-number operator, and $g$ represents the interaction interaction
strength between the measured system and the meter. Here, $\hat{a}^{\dagger}$
and $\hat{a}$ represent the creation and annihilation operators,
respectively. This type of interaction Hamiltonian can be implemented
in optical cavity-QED and solid-state circuit-QED setups \citep{Scully_Zubairy_1997,Gerry_Knight_2004,PhysRevA.69.062320,05461,PhysRevA.84.053846},
and has been employed in various quantum measurement problems \citep{RevModPhys.82.1155,PhysRevLett.114.210801,PhysRevA.99.022114,PhysRevA.106.022619}.

To maintain generality, we assume that the measured system is initially
in a superposition state:
\begin{equation}
\vert\psi_{i}\rangle=\cos\frac{\theta_{i}}{2}\vert g\rangle+e^{i\phi_{i}}\sin\frac{\theta_{i}}{2}\vert e\rangle,\label{eq:intial}
\end{equation}
while the meter initialized in the SPACS defined as:
\begin{equation}
\vert\Phi_{i}\rangle=\gamma\hat{a}^{\dagger}\vert\alpha\rangle,
\end{equation}
where $\gamma=\frac{1}{\sqrt{1+\vert\alpha\vert^{2}}}$ is the normalization
coefficient and $\vert\alpha\rangle$ is a coherent state with complex
amplitude $\alpha=\vert\alpha\vert e^{i\theta}$. As we know if $\alpha\gg1$
the SPACS transferred to coherent state, and in this limit the results
in our paper can reproduce the related claims of Ref. \citep{PhysRevA.106.022619}. 

The unitary operator $\hat{U}$ driven by the interaction Hamiltonian
Eq. (\ref{eq:1}) is given by $\hat{U}=\exp\left(i\lambda\hat{\sigma}_{z}\hat{n}\right)$,
where $\lambda=gt$ is the interaction strength and we set $\hbar=1$
in throughout the work. Under this unitary time evolution operator
the initial composite system state $\vert\psi_{i}\rangle\vert\Phi_{i}\rangle$
transforms into $\hat{U}\vert\psi_{i}\rangle\vert\Phi_{i}\rangle$,
and explicitly expressed as:

\begin{align}
\vert\Phi_{J}\rangle & =\cos\frac{\theta_{i}}{2}\gamma e^{-i\lambda}\vert g\rangle\hat{a}^{\dagger}\vert\xi_{-}\rangle\nonumber \\
 & +e^{i\phi_{i}}\sin\frac{\theta_{i}}{2}\gamma e^{i\lambda}\vert e\rangle\hat{a}^{\dagger}\vert\xi_{+}\rangle,\label{eq:non-posteselected}
\end{align}
where $\xi_{\pm}=\alpha e^{\pm i\lambda}$. As we can see, after time
evolution, the meter and the measured system become entangled. In
the WVA strategy, preselection and postselection are involved in the
measurement process. Here, we assume the postselection state of the
measured system is
\begin{equation}
\vert\psi_{f}\rangle=\cos\frac{\theta_{f}}{2}\vert g\rangle+e^{i\phi_{f}}\sin\frac{\theta_{f}}{2}\vert e\rangle.
\end{equation}
This postselected state has the same form as $\vert\psi_{i}\rangle$,
although the angle parameters $\theta_{i,f}$and $\phi_{i,f}$ are
different. For the above pre- and post-selection state, the corresponding
weak value of measured system observable $\sigma_{z}$ reads as below
\citep{PhysRevLett.60.1351}: 
\begin{align}
\langle\hat{\sigma}_{z}\rangle_{w} & =\frac{\langle\psi_{f}\vert\hat{\sigma}_{z}\vert\psi_{i}\rangle}{\langle\psi_{f}\vert\psi_{i}\rangle}\nonumber \\
 & =\frac{-\cos\frac{\theta_{i}}{2}\cos\frac{\theta_{f}}{2}+e^{i\phi_{0}}\sin\frac{\theta_{i}}{2}\sin\frac{\theta_{f}}{2}}{\cos\frac{\theta_{i}}{2}\cos\frac{\theta_{f}}{2}+e^{i\phi_{0}}\sin\frac{\theta_{i}}{2}\sin\frac{\theta_{f}}{2}},\label{eq:wv}
\end{align}
where $\phi_{0}=\phi_{i}-\phi_{f}$. After selecting the measured
system's postselection state, the final state of the meter becomes:
\begin{align}
\vert\widetilde{\Phi}_{f}\rangle & =\langle\psi_{f}\vert\Phi_{J}\rangle\nonumber \\
 & =\gamma\cos\frac{\theta_{f}}{2}\cos\frac{\theta_{i}}{2}e^{-i\lambda}\hat{a}^{\dagger}\vert\xi_{-}\rangle\nonumber \\
 & +\gamma e^{i\phi_{0}}\sin\frac{\theta_{f}}{2}\sin\frac{\theta_{i}}{2}e^{i\lambda}\hat{a}^{\dagger}\vert\xi_{+}\rangle,\label{eq:w}
\end{align}
 The state $\vert\widetilde{\Phi}_{f}\rangle$ is not normalized.
We define the normalized final meter state as $\vert\Phi_{f}\rangle=\frac{\vert\widetilde{\Phi}_{f}\rangle}{\sqrt{p_{f}}}$,
where $p_{f}=\langle\widetilde{\Phi}_{f}\vert\widetilde{\Phi}_{f}\rangle$
is the probability of successful postselection \citep{PhysRevA.88.042116,PhysRevLett.114.210801}.
The explicit expression for $p_{f}$ reads as 

\begin{align}
p_{f} & =A+\gamma^{2}Be^{-2\vert\alpha\vert^{2}\sin^{2}\lambda}\nonumber \\
 & \times\left[\cos(2\lambda+\phi_{0}+|\alpha|^{2}\sin2\lambda)\right.\nonumber \\
 & +\left.|\alpha|^{2}\cos(4\lambda+\phi_{0}+|\alpha|^{2}\sin2\lambda)\right],
\end{align}
where $\ensuremath{A=\frac{1}{2}(1+\cos\theta_{i}\cos\theta_{f})}$
and $B=\frac{1}{2}\sin\theta_{i}\sin\theta_{f}$. Since the FI provides
a means to quantify the advantages of different strategies, we proceed
to investigate the QFI in both conventional and WVA measurement strategies,
as follow.

\subsection{Quantum Fisher Information with different strategies}

In parameter estimation, the FI quantifies the precision of estimating
an unknown parameter. That is to say, the FI is the maximum amount
of information about the parameter that we can extract from the system.
On the other hand, in quantum metrology, the Cramér-Rao bound (CRB)
can provide a fundamental limit on the precision of parameter estimation.
Specifically, QFI represents the maximum achievable information about
a parameter from measurement outcomes. According to the CRB, the variance
$\Delta^{2}\lambda$ in estimating the parameter $\lambda$ of our
system is fundamentally limited by \citep{PhysRevLett.72.3439,Wiseman_Milburn_2009}:
\begin{equation}
\Delta^{2}\lambda\geq\frac{1}{N\mathcal{F}}.
\end{equation}
where $N$ is the number of measurement trials and $\mathcal{F}$
denotes the FI. 

When no specific measurement strategy is chosen, the QFI encoded in
the state $\vert\Phi\rangle$ is given by:

\begin{equation}
Q_{FI}=4\left(\frac{d\langle\Phi|}{d\lambda}\frac{d|\Phi\rangle}{d\lambda}-\left|\langle\Phi|\frac{d|\Phi\rangle}{d\lambda}\right|^{2}\right),\label{eq:QFI}
\end{equation}
which represents the maximum achievable FI for optimal measurement
of the state $\vert\Phi\rangle$. 

Here, we consider the conventional measurement (cm) without postselection.
For a superposition system state $\vert\psi_{i}\rangle$ as defined
in Eq. (\ref{eq:intial}), the interaction strength $\lambda$ becomes
encoded into the meter state after the interaction between the meter
and the measured system. In non-postselected measurement given by
Eq. (\ref{eq:non-posteselected}), the final state of the meter is
obtained by tracing out the system degrees of freedom as $\rho_{cm}=\text{Tr}_{s}\left(\vert\Phi_{J}\rangle\langle\Phi_{J}\vert\right)$.
As a result, the meter state for the conventional measurement is given
by:

\begin{equation}
\rho_{cm}=\gamma^{2}\left(\cos^{2}\frac{\theta_{i}}{2}\hat{a}^{\dagger}\vert\xi_{-}\rangle\langle\xi_{-}\vert\hat{a}+\sin^{2}\frac{\theta_{i}}{2}\hat{a}^{\dagger}\vert\xi_{+}\rangle\langle\xi_{+}\vert\hat{a}\right),\label{eq:13-1}
\end{equation}
where $\xi_{\pm}=\alpha e^{\pm i\lambda}$. Clearly, $\rho_{cm}$
is a mixed state. The QFI for pure states $\gamma\hat{a}^{\dagger}\vert\xi_{-}\rangle$
and $\gamma\hat{a}^{\dagger}\vert\xi_{+}\rangle$ can be computed
by using Eq. (\ref{eq:QFI}), and both yield the same result:

\begin{equation}
Q_{cm}=4\vert\alpha\vert^{2}\left(2|\alpha|^{2}+4|\alpha|^{4}+|\alpha|^{6}-\vert\alpha\vert^{2}\left(2|\alpha|^{2}+|\alpha|^{4}\right)^{2}\right).\label{eq:Qcm-1}
\end{equation}
Since both components of the mixed meter state yield identical QFI
$Q_{cm}$, the total QFI of the final meter state $\rho_{cm}$ in
the conventional measurement is also given by $Q_{cm}$. It is important
to note that this quantity remains constant for a fixed $\alpha$,
and does not depend on the interaction strength $\lambda$. As a result,
the estimation precision cannot be improved by tuning interaction
strength $\lambda$, and is instead solely determined by the fixed
meter state parameter $\alpha$.

When the WVA strategy is employed, the effective QFI, denoted as $F_{tot}=p_{f}Q_{FI}$,
quantifies the maximum amount of QFI available in the WVA measurement.
Explicitly, in our system the final meter state after postselection
is $\vert\Phi_{f}\rangle$, the WVA-QFI is given by 

\begin{widetext}

\begin{align}
F_{tot} & =p_{f}Q_{FI}\nonumber \\
 & =4\left[\gamma^{2}A\left(1+3\langle\hat{n}\rangle+3\langle\hat{n}^{2}\rangle+\langle\hat{n}^{3}\rangle\right)-\frac{\gamma^{4}C^{2}\left(1+2\langle\hat{n}\rangle+\langle\hat{n}^{2}\rangle\right)^{2}}{p_{f}}\right.\nonumber \\
 & -\gamma^{2}Be^{-|\alpha|^{2}}\sum_{n=0}^{\infty}\frac{\left(|\alpha|^{2}\right)^{n}}{n!}\left(1+3n+3n^{2}+n^{3}\right)\cos\left(2\lambda+2n\lambda+\phi_{0}\right)\label{eq:11}\\
 & -\left.\frac{B^{2}\gamma^{4}\left[e^{-|\alpha|^{2}}\sum_{n=0}^{\infty}\frac{|\alpha|^{2}}{n!}\left(1+2n+n^{2}\right)\sin\left(2\lambda+2n\lambda+\phi_{0}\right)\right]^{2}}{p_{f}}\right],\nonumber 
\end{align}
\end{widetext}where $\ensuremath{C=\frac{1}{2}(\cos\theta_{i}+\cos\theta_{f})}$.
In Eq. (\ref{eq:11}), the expectation values are explicitly given
by $\langle\hat{n}\rangle=|\alpha|^{2}$, $\langle\hat{n}^{2}\rangle=|\alpha|^{2}+|\alpha|^{4}$,
and $\langle\hat{n}^{3}\rangle=|\alpha|^{2}+3|\alpha|^{4}+|\alpha|^{6}$. 

The expression for $F_{tot}$ becomes complicated with varying the
system parameters. Therefore, in the next section, we numerically
investigate its behavior under different parameter regimes and compare
it with alternative measurement strategies.

Therefore, in the next subsection, we numerically investigate its
behavior under different parameter regimes and compare it with alternative
measurement methods. There are alternative methods to estimate parameters
in quantum measurements with WVA strategy. We specifically focus on
two widely used quantum measurement techniques photon-number measurement
and field-quadrature measurement—and illustrate the advantages of
the WVA strategy for accurately extracting information about the interaction
strength $\lambda$.

\subsection{Photon-Number measurement}

The final meter is characterized by the state $\vert\Phi_{f}\rangle$,
which evolves from the SPACS and has a corresponding photon-number
distribution. In this subsection, we calculate the photon-number measurement
assisted FI for estimating the parameter $\lambda$, encoded in the
meter state $\vert\Phi_{f}\rangle$. The photon-number probability
distribution for the final meter state can be expressed as:
\begin{align}
P_{f}(n) & =\vert\langle n\vert\Phi_{f}\rangle\vert^{2}=\frac{1}{p_{f}}\vert\langle n\vert\widetilde{\Phi}_{f}\rangle\vert^{2}\nonumber \\
 & =\frac{\gamma^{2}}{p_{f}}\frac{n|\alpha|^{2n-2}e^{-|\alpha|^{2}}}{(n-1)!}[A+B\cos(2n\lambda+\phi_{0})].
\end{align}
This expression shows that photon-number distribution encodes information
about the interaction strength $\lambda$. Specifically, for the given
values of $\alpha$, $\theta_{f(i)}$, $\phi_{f(i)}$, and other system
parameters, the photon-number distribution function $P_{f}(n)$ corresponds
precisely to the conditional probability $P(n\vert\lambda)$, where
$\lambda$ is the unknown parameter of interest. The goal is thus
to estimate the true value of $\lambda$ as accurately as possible
using measurement outcomes $n$. This estimation is carried out by
an appropriate estimator function $\lambda(x)$, whose uncertainty
is fundamentally bounded by the CRB \citep{Fisher1925,RN21} as mentioned
above context. Here, note the corresponding FI with $F_{f}^{(ph)}$
to estimate the interaction strength parameter $\lambda$ from the
photon-number measurements in a postselected weak measurement scheme
and it is defined as
\begin{equation}
F_{f}^{(ph)}=\sum_{n}\frac{1}{P_{f}(n)}\left(\frac{\partial P_{f}(n)}{\partial\lambda}\right)^{2}.\label{eq:11-1}
\end{equation}
This FI also can quantify the minimal achievable root mean square
error in estimating $\lambda$ and sets a fundamental limit on the
precision of our measurement scheme. However, since our measurement
scheme includes a postselection process after the unitary evolution
of the composite system, we must account for the success probability
$p_{f}$ in the FI. Hence, we introduce the effective FI, defined
as $p_{f}F_{f}^{(ph)}$, to quantify the precision attainable through
photon-number measurements.

\begin{figure}
\includegraphics[width=4cm]{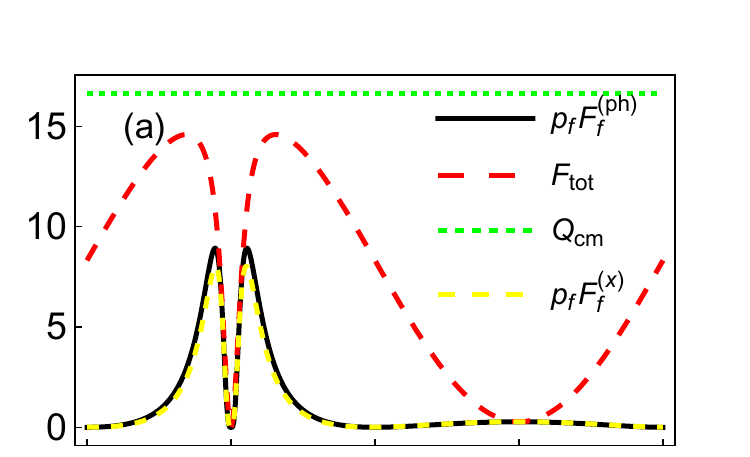}\includegraphics[width=4cm]{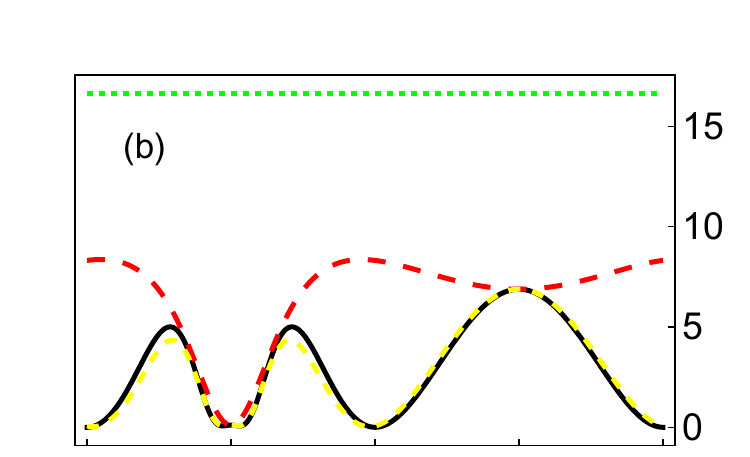}

\includegraphics[width=4cm]{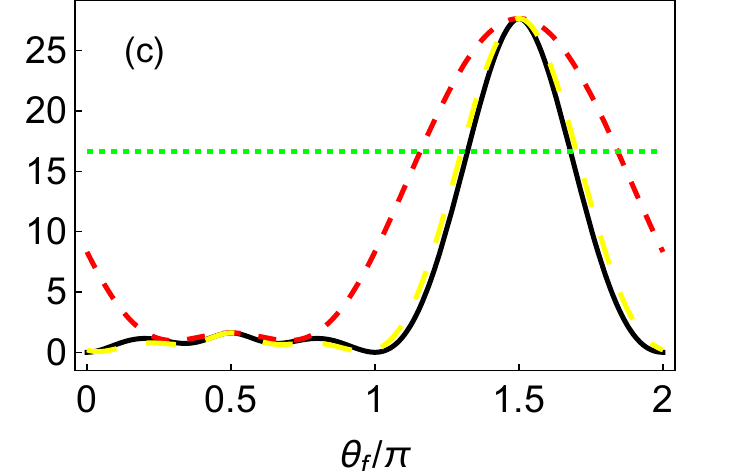}\includegraphics[width=4cm]{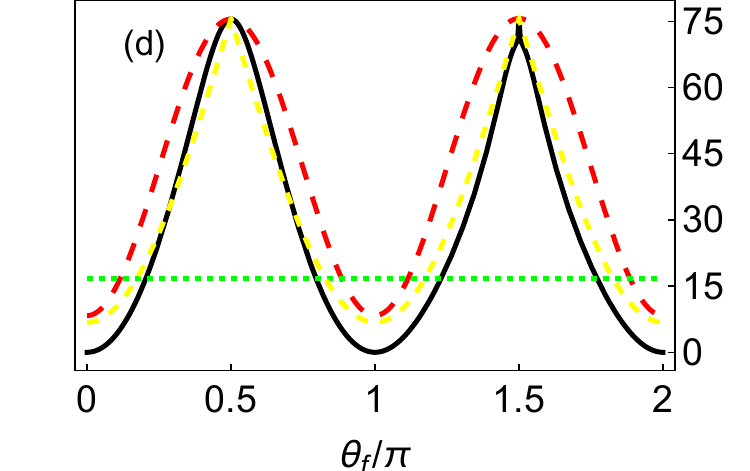}

\caption{\label{fig:WVA-FI and QFI}The \textcolor{blue}{WVA}-FI $p_{f}F_{f}^{(ph,x)}$associated
with the photon-number (black solid curve) and $x$-quadrature (yellow
dashed curve) measurements are compared with the WVA-QFI $F_{tot}$
(red dashed curve) and the QFI $Q_{cm}$ (green dotted line) of conventional
measurement.\textcolor{blue}{{} WVA-FI $p_{f}F_{f}^{(ph,x)}$ and WVA-QFI
$F_{tot}$ involve postselection and are functions of $\theta_{f}$
with a period of $2\pi$}. Panels show comparisons for different interaction
strengths: (a) $\lambda=0.01$, (b) $\lambda=0.05$, (c) $\lambda=0.1$,
(d) $\lambda=1$. The other parameters are $\theta_{i}=\frac{\pi}{2}$
, $\phi_{0}=\pi$ , and $\alpha=2$.}
\end{figure}

To clearly illustrate the behavior of the effective FI $p_{f}F_{f}^{(ph)}$,
we plot it as a function of the postselection angle $\theta_{f}$
for different interaction strength parameters $\lambda$ in Fig. \ref{fig:WVA-FI and QFI}.
The plotting parameters are the same as those in Ref. \citep{PhysRevA.106.022619},
allowing for a direct comparison between SPACS and coherent states
as meter under the same postselected measurement scheme. As shown
in Fig. \ref{fig:WVA-FI and QFI}, for the optimal choice of $\theta_{f}$,
the effective FI $p_{f}F_{f}^{(ph)}$can attain the WVA-QFI value
$F_{tot}$. For other values of $\theta_{f}$, the effective FI is
always lower than the WVA-QFI (see the black solid curves in Fig.
\ref{fig:WVA-FI and QFI}).

\subsection{Field-Quadrature Measurement}

Next, we turn to the field-quadrature measurement to extract information
about the parameter $\lambda$ encoded in the SPACS. This measurement
can be implemented using a homodyne detection scheme, in which the
meter field is mixed with a strong local oscillator serving as a reference
field with a well-defined phase $\vartheta$ \citep{Wiseman_Milburn_2009}.
The quadrature operator measured in this scheme is given by:

\begin{equation}
\hat{x}=(\hat{a}e^{-i\vartheta}+\hat{a}^{\dagger}e^{i\vartheta})/\sqrt{2},
\end{equation}
with the corresponding conjugate quadrature operator given by:

\begin{equation}
\hat{p}=-i(\hat{a}e^{-i\vartheta}-\hat{a}^{\dagger}e^{i\vartheta})/\sqrt{2},
\end{equation}
which satisfy the commutation relation $[\hat{x},\hat{p}]=i$. For
simplicity, when the local oscillator phase is set to zero ($\vartheta=0$),
the quadratures reduce to $\hat{x}=(\hat{a}+\hat{a}^{\dagger})/\sqrt{2}$
and $\hat{p}=-i(\hat{a}-\hat{a}^{\dagger})/\sqrt{2}$.

To proceed with the calculation, we evaluate the wave function of
the meter state in the coordinate representation as
\begin{align}
\langle x\vert\gamma e^{-i\lambda}\hat{a}^{\dagger}\vert\xi_{-}\rangle & =\gamma e^{-i\lambda}e^{-|\alpha|/2}e^{\frac{x^{2}}{2}}\left(\frac{1}{\pi}\right)^{\frac{1}{4}}\\
 & \times\left(\sqrt{2}x-\alpha e^{-i\lambda}\right)\exp\left[-\left(\frac{\alpha e^{-i\lambda}}{\sqrt{2}}-x\right)^{2}\right].\nonumber 
\end{align}
After postselection, the meter state becomes a superposition of $\gamma e^{-i\lambda}\hat{a}^{\dagger}\vert\alpha e^{-i\lambda}\rangle$
and $\gamma e^{i\lambda}\hat{a}^{\dagger}\vert\alpha e^{i\lambda}\rangle$,
as shown in Eq. (\ref{eq:w}). The probability distribution function
of the $x$-quadrature measurement then reads

\begin{widetext}

\begin{align}
P_{f}(x) & =\vert\langle x\vert\Phi_{f}\rangle\vert^{2}=\frac{1}{p_{f}}\vert\langle x\vert\widetilde{\Phi}_{f}\rangle\vert^{2}\nonumber \\
 & =\frac{\gamma^{2}e^{-|\alpha|^{2}}\left(\frac{1}{\pi}\right)^{\frac{1}{2}}}{p_{f}}A\left(2x^{2}-2\sqrt{2}x\alpha\cos\lambda+|\alpha|^{2}\right)e^{-\alpha^{2}\cos2\lambda+2\sqrt{2}x\alpha\cos\lambda-x^{2}}\nonumber \\
 & +\frac{\gamma^{2}e^{-|\alpha|^{2}}\left(\frac{1}{\pi}\right)^{\frac{1}{2}}}{p_{f}}B\mathrm{Re}\left[e^{2i\lambda}e^{-i\phi_{0}}\left(2x^{2}-2\sqrt{2}x\alpha e^{i\lambda}+|\alpha|^{2}e^{2i\lambda}\right)e^{-\alpha^{2}e^{2i\lambda}+2\sqrt{2}x\alpha e^{i\lambda}-x^{2}}\right].\label{eq:17}
\end{align}

\end{widetext}From this expression, we see that the probability distribution
function $P_{f}(x)$ contains the information about the parameter
$\lambda$ and is also expressed as $P_{f}(x)=P(x\vert\lambda)$.
We obtain the FI associated with the probability distribution $P_{f}(x)$
by extending Eq. (\ref{eq:11-1}) from summation to integration, as
$x$ is a continuous variable. Thus, the FI corresponding to the parameter
$\lambda$ encoded in $P_{f}(x)$ is given by:
\begin{equation}
F_{f}^{(x)}=\int dx\frac{1}{P_{f}(x)}\left(\frac{\partial P_{f}(x)}{\partial\lambda}\right)^{2}.\label{eq:18}
\end{equation}

Although the integral expression for the FI is well-defined, its direct
evaluation is complicated. To simplify this issue, approximate the
integral in Eq. (\ref{eq:18}) using a summation over a finite range
$-10\leq x\leq10$, which ensures convergence of the integrand. The
Eq. (\ref{eq:18}) being integrated has already converged in this
range, so the summation method provides a good approximation. Therefore,
we divide the interval into 2000 subintervals and compute the sum
of the areas of these subintervals using the rectangular area for
numerical integration as follow:
\begin{equation}
F_{f}^{(x)}\approx\sum_{y=-1000}^{1000}\frac{1}{100P_{f}(y)}\left(\frac{\partial P_{f}(y)}{\partial\lambda}\right)^{2},\label{eq:19}
\end{equation}
where $y=100x$. In this way, we can quantify the Fisher information
$p_{f}F_{f}^{(x)}$ by including the postselection effect. The numerical
results are displayed in Fig. \ref{fig:WVA-FI and QFI}.

\subsection{Comparison different methods}

In above contexts, we have defined and calculated four different forms
of FI to quantify the estimation precision of the parameter $\lambda$.
To clearly illustrate the effectiveness of these measurement strategies,
Fig. \ref{fig:WVA-FI and QFI} shows a direct comparison among them.
Specifically, the FI corresponding to photon number and field-quadrature
measurements, both with postselection, are compared with the WVA-QFI
($F_{tot}$) and the conventional measurement QFI ($Q_{cm}$) in the
same plots. Several conclusions can be drawn from Fig. \ref{fig:WVA-FI and QFI}:

Photon-number and field-quadrature measurements yield comparable precision
in the WVA strategy. Generally, the values of $p_{f}F_{f}^{(n)}$
and $p_{f}F_{f}^{(x)}$ are lower than the WVA-QFI $F_{tot}$, except
at specific optimal angles $\theta_{f}$, where they coincide with
$F_{tot}$. Notably, the WVA-based methods outperform conventional
measurements at optimal postselection angles, especially when $\lambda\geq0.1$.
Utilizing SPACS as the meter consistently provides enhanced measurement
sensitivity compared to conventional measurement schemes. Moreover,
as the FI increases, the precision of estimating the interaction strength
$\lambda$ improves. 

In previous work \citep{PhysRevA.106.022619}, it was demonstrated
that using coherent states as meters in WVA schemes achieves higher
FI than conventional measurements, but only under larger interaction
strengths that lie beyond the weak measurement regime, namely, $\lambda\ge1$.
However, as shown by our analysis, the SPACS-based meter is more sensitive
than the coherent-state meter in parameter estimation due to the inherent
nonclassical features of SPACS. This enhanced sensitivity arises directly
from the nonclassical nature of the SPACS, consistent with findings
reported in Refs. \citep{PhysRevLett.105.010403,RN17}.

\section{\label{sec:4}State distance and  SNR}

In quantum metrology, the state distance and the SNR are essential
metrics for evaluating the effectiveness of measurement strategies.
The state distance, often quantified by fidelity, characterizes how
the quantum state of a system evolves under a specific interaction
and provides insight into the closeness between the initial and final
states. On the other hand, the SNR quantifies how effectively a measurement
can extract information from a noisy quantum system, with higher SNR
indicating greater measurement accuracy. In this section, we analyze
the state distance between meter states for arbitrary interaction
strength $\lambda$, and we discuss the SNR in postselected measurements
compared to conventional (non-postselected) measurements.

\subsection{State Distance}

The fidelity function $F=\vert\langle\phi_{0}\vert\phi_{t}\rangle\vert^{2}$
can characterize how close a state $\vert\phi_{0}\rangle$ is to the
target state $\vert\psi_{t}\rangle$. It is equal to one if the two
states are identical, while it is zero when the two ate orthogonal
to each other. In our scheme, the fidelity between the initial meter
state $\vert\Phi_{i}\rangle$ and postmeasurement state $\vert\Phi_{f}\rangle$
is given by

\begin{align}
F & =\frac{\gamma^{4}}{p_{f}}|\cos\frac{\theta_{f}}{2}\cos\frac{\theta_{i}}{2}(1+|\alpha|^{2}e^{i\lambda})e^{-|\alpha|^{2}+|\alpha|^{2}e^{i\lambda}+i\lambda}\nonumber \\
 & +\sin\frac{\theta_{f}}{2}\sin\frac{\theta_{i}}{2}(1+|\alpha|^{2}e^{-i\lambda})e^{-|\alpha|^{2}+|\alpha|^{2}e^{-i\lambda}-i(\phi_{0}+\lambda)}|^{2}.\label{eq:22}
\end{align}

\begin{figure}
\includegraphics[width=8cm]{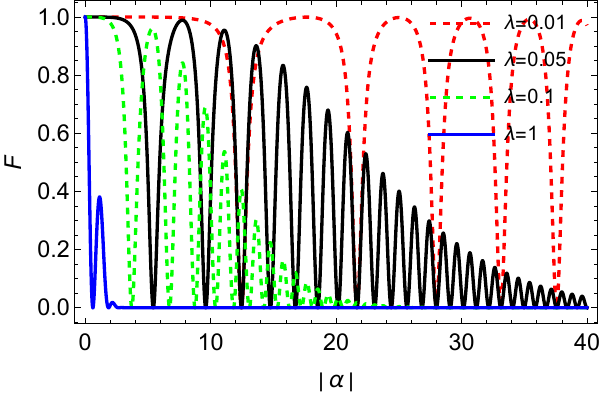}

\caption{\label{fig:The-state-distance}The state distance of the meter state
as a function of $\vert\alpha\vert$ in postselected measurements
for different interaction strengths $\lambda$. The curves correspond
to $\lambda=0.01$ (red dashed curve), $\lambda=0.05$ (black solid
curve), $\lambda=0.1$ (green dashed curve), and $\lambda=1$ (blue
solid curve). The other parameters are $\theta_{i}=\frac{\pi}{2}$
, $\phi_{0}=\pi$ , and $\theta_{f}=\frac{3\pi}{2}$}
\end{figure}

To provide a clearer analysis of the fidelity behavior, we plot it
as a function of the state parameter $|\alpha|$, as shown in Fig.
\ref{fig:The-state-distance}. From Fig. \ref{fig:The-state-distance},
we can observe that the state distance exhibits periodic oscillations
as a function of $|\alpha|$ for different interaction strengths $\lambda$.
The amplitude and period of these oscillations decrease as $|\alpha|$
increases and become more pronounced for larger values of $\lambda$.
For instance, when the interaction strength is $\lambda=0.1$, the
fidelity reaches zero at $|\alpha|=25$, whereas for $\lambda=1$,
it already drops to zero at $|\alpha|=2$. 

The fidelity $F$ directly reflects how strongly the interaction strength
parameter $\lambda$ perturbs the meter state. A lower fidelity implies
greater distinguish ability between the initial and final states with
postselected measurement, which correlates with higher sensitivity
to $\lambda$. The periodic oscillations of $F$ with respect to state
parameter $\vert\alpha\vert$ arise from quantum interference between
the two terms in $\vert\Phi_{f}\rangle$, which depend on the interaction
phase $e^{\pm i\lambda}$. In Eq. (\ref{eq:22}) the term $e^{-|\alpha|^{2}+|\alpha|^{2}e^{\pm i\lambda}}=e^{|\alpha|^{2}\left(\cos\lambda-1\right)}e^{\pm i|\alpha|^{2}\sin\lambda}$
shows that the first exponential factor $e^{|\alpha|^{2}\left(\cos\lambda-1\right)}$
modulates the amplitude, while the second exponential factor governs
the oscillatory behavior. From Fig. \ref{fig:The-state-distance},
we can see that for lager interaction interaction strength parameter
$\lambda=1$, the state distance dramatically decrease as $\vert\alpha\vert\gtrsim2$,
which would limit its piratical applications where a low average photon
number. However, in the weak interaction interaction regime, the high
fidelity region is broadened, and it become possible to optimize the
fidelity over different oscillation periods.

A high fidelity implies that the final postselected meter state remains
highly similar to the initial meter state, its mean the meter can
be reused for further measurement. Therefore, exploiting weak measurements
in precision metrology is meaningful. As shown in Fig. \ref{fig:The-state-distance},
PACS as meter keep high fidelity in $\vert\alpha\vert=2$ with $\lambda=0.1$,
indicating that the meter can be recycled while simultaneously realizing
a higher effective FI than conventional measurement schemes, as shown
in Fig .\ref{fig:WVA-FI and QFI} (c).

\subsection{SNR}

We analyze the SNR between postselected and conventional measurements
to evaluate the effectiveness of SPACS-based postselected measurements
in improving measurement precision. The ratio of the SNR between postselected
and conventional measurements is defined as:
\begin{equation}
\eta=\frac{S_{x}^{p}}{S_{x}^{n}},
\end{equation}
where $S_{x}^{p}$ and $S_{x}^{n}$ correspond to the SNR with and
without postselection, respectively. The position operator $\hat{x}=(\hat{a}+\hat{a}^{\dagger})/\sqrt{2}$
is used to characterize the measurement process. As mentioned in Sec.
\ref{sec:2}, in the absence of postselection, the system and meter
undergo coupled evolution via the interaction Hamiltonian, as described
by the state in Eq. (\ref{eq:non-posteselected}).

The SNR for conventional measurements is defined as:

\begin{equation}
S_{x}^{n}=\frac{\sqrt{N}|\delta x|}{\sqrt{\langle\hat{x}^{2}\rangle_{J}-\langle\hat{x}\rangle_{J}^{2}}}=\frac{\sqrt{N}|\langle\hat{x}\rangle_{J}-\langle\hat{x}\rangle_{i}|}{\sqrt{\langle\hat{x}^{2}\rangle_{J}-\langle\hat{x}\rangle_{J}^{2}}},
\end{equation}
where $N$ is the total number of measurements, $\left\langle \bullet\right\rangle _{i}$
and $\left\langle \bullet\right\rangle _{J}$ represent the expectation
values of related quantities under the initial meter state $\vert\Phi_{i}\rangle$
and the composite state after the interaction $\vert\Phi_{J}\rangle$,
respectively. For postselected measurements, the SNR is given by \citep{GUO2021104868}:

\begin{equation}
S_{x}^{p}=\frac{\sqrt{p_{f}N}|\delta x|}{\sqrt{\langle\hat{x}^{2}\rangle_{f}-\langle\hat{x}\rangle_{f}^{2}}}=\frac{\sqrt{p_{f}N}|\langle\hat{x}\rangle_{f}-\langle\hat{x}\rangle_{i}|}{\sqrt{\langle\hat{x}^{2}\rangle_{f}-\langle\hat{x}\rangle_{f}^{2}}}.
\end{equation}
where $\left\langle \bullet\right\rangle _{f}$ represents the expectation
value under the meter state $\vert\Phi_{f}\rangle$ with the postselection.

\begin{figure}
\includegraphics[width=8cm]{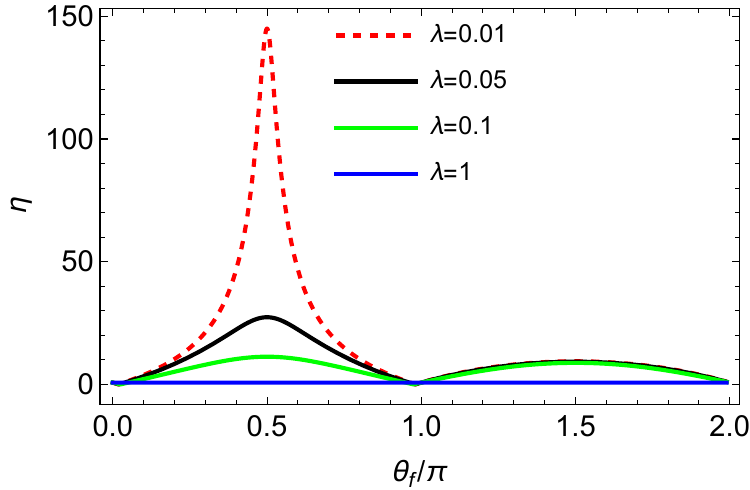}

\caption{\label{fig:the-signal-to-noise-ratio} The ratio of SNR between postselected
and conventional measurements as function of the postselected angle
$\theta_{f}$ for different interaction strengths $\lambda$. Here,
$\lambda=0.01$ (red dashed curve) , $\lambda=0.05$ (black solid
curve), $\lambda=0.1$ (green dashed curve), $\lambda=1$ (blue solid
curve). The other parameters are the same as Fig. \ref{fig:WVA-FI and QFI}.}
\end{figure}

The expectation values of the position operator can be calculated
using the expressions for $\vert\Phi_{i}\rangle$, $\vert\Phi_{J}\rangle$
and $\vert\Phi_{f}\rangle$. In Fig. \ref{fig:the-signal-to-noise-ratio},
we present the ration $\eta$ as a function of the postselection angle
$\theta_{f}$ for different interaction strengths $\lambda$. A ratio
$\eta>1$ implies that the measurement with postselection extracts
more information than the conventional measurements.

As shown in Fig. \ref{fig:the-signal-to-noise-ratio}, the ratio $\eta$
exhibits periodic behavior dependence on the postselection angle,
and we plot it over a single period interval. For larger interaction
strength (e.g., $\lambda=1$), the ratio $\eta\approx0.7$, indicating
the SNR for postselected measurement is lower compere to conventional
measurement in this condition. However, the ratio $\eta$ is higher
for small interaction strengths, demonstrating that WVA is more effective
under weak measurement. Overall, the ratio of SNR decreases as the
interaction strength $\lambda$ increases. By choosing an optimal
postselection angle (e.g., $\theta_{f}\approx\frac{\pi}{2}$, where
$\eta$ is approximately $144.4$), postselected weak measurements
can significantly enhance the signal while suppressing technical noise,
offering clear advantages over conventional measurement scheme. In
this case, the weak value becomes anomalously large in Eq. (\ref{eq:wv}).
This result further underscores the utility of postselected weak measurements
in precision metrology applications \citep{2020A}. 

\section{\label{sec:5}Weak Measurement Proposal Based on Photon Statistics}

This section presents an alternative weak measurement proposal based
on the photon statistics of SPACS. The goal is to explore how the
real and imaginary components of the weak values of the measured system's
observable can be extracted using photon statistics as the meter in
a specific class of weak measurements. 

Firstly, we briefly retrospect the weak value. In postselected weak
measurement theory only the low-order Taylor expansion for the evolution
operator of the measured system and pointer is considered \citep{PhysRevLett.60.1351}
so that the measured system almost not disturbed after single trial.
Even if we didn't obtain enough information in single trial, we can
repeat the undisturbed weak measurement many times and finally can
achieve desired results statistically. Furthermore, the normal expectation
and conditional expectation values of a system observable always can
be written in terms of weak value, and this connection play important
role to our understanding of weak-to-strong measurement transition
\citep{TUREK2023128663}.

In the original work of Aharonov et al., the interaction between the
measured system and the meter is described by the Hamiltonian $\hat{H}_{\mathrm{int}}=g\hat{A}\otimes\hat{p}$,
where $\hat{A}$ is the observable of the measured system, $\hat{p}$
is the meter's momentum operator, and $g$ represents the weak interaction
strength. In the context of pre-selection and postselection with initial
and final states $\vert\psi_{i}\rangle$ and $\vert\psi_{f}\rangle$,
respectively, the weak value $A_{w}$ of the measured system's observable
is defined as \citep{PhysRevLett.60.1351}:

\begin{equation}
A_{w}=\frac{\langle\psi_{f}|\hat{A}|\psi_{i}\rangle}{\langle\psi_{f}|\psi_{i}\rangle}.\label{eq:weak value}
\end{equation}
In general, the weak value manifests as a complex quantity, with its
real and imaginary components corresponding to different physical
quantities. In measurement processes, the real part, $\mathrm{Re}(A_{w})$,
is associated with the shift in the meter’s position $x$, while the
imaginary part, $\mathrm{Im}(A_{w})$, corresponds to the shift in
the meter's momentum $p$ \citep{LUNDEEN2005337}. This duality arises
from the canonical commutator $[\hat{x},\hat{p}]=i$, which ensures
that the real and imaginary components of $A_{w}$ encode complementary
information about the system's interaction with the meter. 

Here, we introduce a weak measurement formalism applies to the photon
number-phase variables in quantum optics. We have to mention that
our measurement scheme below still belong to a special case of original
postselected weak measurement theory proposed by Aharonov and his
coworkers \citep{PhysRevLett.60.1351}. 

In a photonic system, the conjugate pair consisting of the photon-number
operator $\hat{n}=\hat{a}^{\dagger}\hat{a}$ and the phase operator
$\hat{\varphi}$ is expected to satisfy the commutation relation 
\begin{equation}
[\hat{n},\hat{\varphi}]=i.\label{eq:31}
\end{equation}
However, this commutation relation lack rigorous proof since defining
a phase operator in physical systems is inherently challenging. Despite
the absence of a universally accepted phase operator $\hat{\varphi}$,
various studies suggest the validity of the above commutation relation
\citep{Barnett01011989,RN15}. By analogy with the $\hat{x}$–$\hat{p}$
case, in postselected weak measurements involving the conjugate pair
of operators $\hat{n}$ and $\hat{\varphi}$, the real and imaginary
parts of the weak value should correspond to these conjugate variables.

We begin our postselected weak measurement with the with interaction
Hamiltonian given Eq. (\ref{eq:1}). We assume that the photon statistics
of SPACS as a meter and $\hat{\sigma}_{z}$ is measured observable.
Comparing the interaction Hamiltonian of our scheme in Eq. (\ref{eq:1})
with standard von Neumann measurement Hamiltonian $\hat{H}_{\mathrm{int}}=g\hat{A}\otimes\hat{p}$,
it becomes clear that $\hat{A}$ and $\hat{p}$ correspond to $\hat{\sigma}_{z}$
and the number operator $\hat{n}=\hat{a}^{\dagger}\hat{a}$, respectively.
Along with the standard measurement procedures of postselected weak
measurement by considering pre- and post-selected measured system
states $\vert\psi_{i}\rangle$ and $\vert\psi_{f}\rangle$ and initial
meter state $\vert\Phi_{i}\rangle$, the state of the meter evolves
to $\vert\widetilde{\Phi}_{w}\rangle=\langle\psi_{f}\vert\exp\left(-i\lambda\hat{\sigma}_{z}\hat{n}\right)\vert\psi_{i}\rangle\vert\Phi_{i}\rangle$
(unnormalized), where $\lambda=gt$. When $\lambda\langle\hat{\sigma}_{z}\rangle_{w}\ll1$,
the final meter sate with postselected weak measurement can be derive
as follow: 

\begin{equation}
\vert\Phi_{w}\rangle\approx\kappa a^{\dagger}\vert\beta\rangle.\label{eq:w-p}
\end{equation}
where, $\kappa=\frac{1}{\sqrt{1+\vert\beta\vert^{2}}}$ and $\beta=\alpha e^{-i\lambda\langle\hat{\sigma}_{z}\rangle_{w}}$,
and $\langle\hat{\sigma}_{z}\rangle_{w}$ is weak value of measured
system observable $\hat{\sigma}_{x}$ for pre- and postselected states
$\vert\psi_{i}\rangle$ and $\vert\psi_{f}\rangle$ defined in Sec.
\ref{sec:2}. It is important to note that in the above derivation,
we used the approximation $1-i\lambda\langle\hat{\sigma}_{z}\rangle_{w}\approx e^{-i\lambda\langle\hat{\sigma}_{z}\rangle_{w}}$
for $\lambda\langle\hat{\sigma}_{z}\rangle_{w}\ll1$. The average
photon number associated with this final meter state depends on the
weak value. In the weak measurement regime, the shift in the average
photon number is given by 
\begin{align}
\delta n & =\langle\Phi_{w}\vert\hat{n}\vert\Phi_{w}\rangle-\langle\Phi_{i}\vert\hat{n}\vert\Phi_{i}\rangle\nonumber \\
 & \approx\frac{2\lambda\vert\alpha\vert^{2}\gamma^{2}\mathrm{Im}[\langle\hat{\sigma}_{z}\rangle_{w}]}{2\lambda|\alpha|^{2}\mathrm{Im}[\langle\hat{\sigma}_{z}\rangle_{w}]+|\alpha|^{2}+1}\left(\vert\alpha\vert^{4}+2\vert\alpha\vert^{2}+2\right).\label{eq:30}
\end{align}
As expected, the $\delta n$ is proportional to the imaginary part
of the weak value and reduces to coherent state pointer-based result,
$2\lambda\vert\alpha\vert^{2}\mathrm{Im}[\langle\hat{\sigma}_{z}\rangle_{w}]$,
if $\vert\alpha\vert$$\gg1$. 

\begin{figure}
\includegraphics[width=8cm]{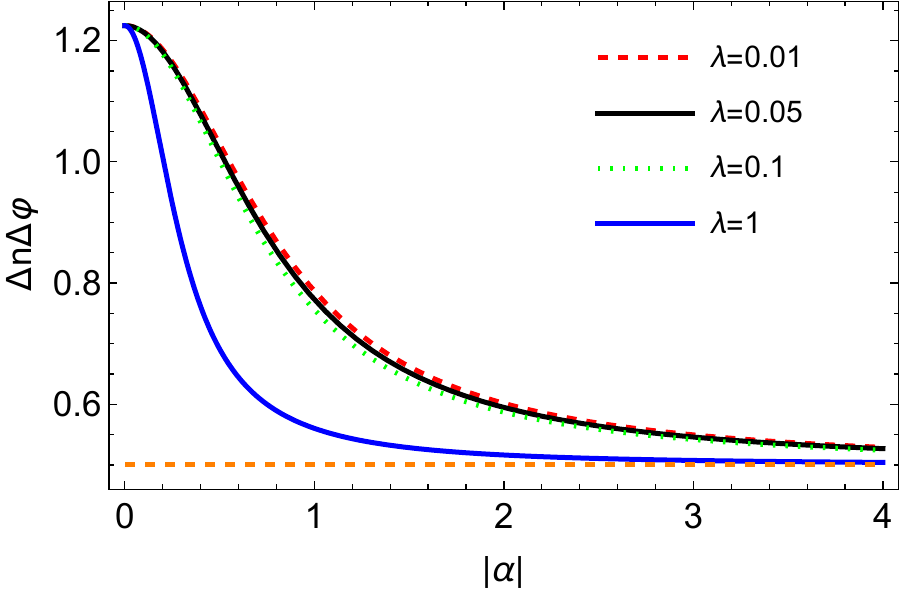}

\caption{\label{fig:4}The number-phase uncertainty product $\Delta n\Delta\varphi$
as function of $\vert\alpha\vert$ for different interaction strengths
$\lambda.$ Here, we take the weak value $\langle\hat{\sigma}_{z}\rangle_{w}=1+i$,
with $\lambda=0.01$ (green solid curve) , $\lambda=0.05$ (black
solid curve), $\lambda=0.1$ (green dashed curve), $\lambda=1$ (blue
solid curve). The orange dashed line represents the number-phase uncertainty
product $\Delta n\Delta\varphi$ for the coherent state. }

\end{figure}

Next, we investigate the readout method for the real part of the weak
value in our weak measurement scheme. We consider the discrete basis
of phase eigenvectors in the form: $\vert\varphi\rangle=\frac{1}{\sqrt{S+1}}\sum_{n=0}^{S}e^{in\varphi}\vert n\rangle,$as
proposed by Pegg and Barnett \citep{Gerry_Knight_2004,Barnett01011989,RN15}.
For finite $S$ the phase state $\vert\varphi\rangle$ is discrete,
with the phase parameter taking the discrete values $\varphi_{m}=\varphi_{0}+\frac{2\pi m}{S+1},\quad m=0,1,...,S.$
As $S\rightarrow\infty$, the phase state $\vert\varphi\rangle$ transferred
from a discrete state to a continuous state
\begin{equation}
\vert\varphi\rangle=\sum_{n=0}^{\infty}e^{in\varphi}\vert n\rangle.
\end{equation}
The phase $\varphi$ can then take any continuous value in the interval
$[0,2\pi)$. The completeness relation for the continuous phase states
is given by \citep{Gerry_Knight_2004} 
\begin{equation}
\frac{1}{2\pi}\int\vert\varphi\rangle\langle\varphi\vert d\phi=1.
\end{equation}
From Eq. (\ref{eq:w-p}), we see that the final pointer state $\vert\Phi_{w}\rangle$
remains a SPACS. Using the phase state $\vert\varphi\rangle$, we
define the phase distribution of an arbitrary state $\vert\psi\rangle$
as 
\begin{equation}
P\left(\varphi\right)\equiv\frac{1}{2\pi}\vert\langle\varphi\vert\psi\rangle\vert^{2}.\label{eq:32}
\end{equation}
This phase distribution satisfies the normalization condition $\int_{0}^{2\pi}P(\varphi)d\varphi=1$,
provided that the state $\vert\psi\rangle$ is normalized. One of
the key applications of the phase distribution function $P(\varphi)$
is that it allows us to compute the expectation value of any function
of $\varphi$, denoted as $f(\varphi)$, using 
\begin{equation}
\langle f(\varphi)\rangle=\int_{0}^{2\pi}f(\varphi)P(\varphi)d\varphi.\label{eq:34}
\end{equation}
For our scheme, the phase distributions for the initial and final
pointer states, $\vert\Phi_{i}\rangle$ and $\vert\Phi_{w}\rangle$,
are given by: 

\begin{subequations}
\begin{align}
P_{i}(\varphi) & =\frac{1}{2\pi}\vert\langle\varphi\vert\Phi_{i}\rangle\vert^{2}\nonumber \\
 & \approx\gamma^{2}\sqrt{\frac{2}{\pi}}\vert\alpha\vert^{3}\left[4\left(\theta-\varphi\right)^{2}+1\right]e^{-2\vert\alpha\vert^{2}\left(\theta-\varphi\right)^{2}}\label{eq:35}
\end{align}
and 
\begin{align}
P_{w}(\varphi) & =\frac{1}{2\pi}\vert\langle\varphi\vert\Phi_{w}\rangle\vert^{2}\nonumber \\
 & \approx\kappa^{2}\sqrt{\frac{2}{\pi}}\vert\beta\vert^{3}\left[4\text{\ensuremath{\left(\theta-\lambda\mathrm{Re}[\langle\hat{\sigma}_{z}\rangle_{w}]-\varphi\right)}}^{2}+1\right]\nonumber \\
 & \times e^{-2\vert\beta\vert^{2}\text{\ensuremath{\left(\theta-\lambda\mathrm{Re}[\langle\hat{\sigma}_{z}\rangle_{w}]-\varphi\right)}}^{2}}.\label{eq:37}
\end{align}
\end{subequations} In deriving these expressions, we used the fact
that for large $\vert\alpha\vert^{2}$, the Poisson distribution can
be approximated by a Gaussian distribution, i.e., 
\begin{equation}
\begin{aligned}\frac{|\alpha|^{2n}}{n!}\mathrm{e}^{-|\alpha|^{2}} & \approx\left(2\pi|\alpha|^{2}\right)^{-1/2}\exp\left[-\frac{\left(n-|\alpha|^{2}\right)^{2}}{2|\alpha|^{2}}\right].\end{aligned}
\end{equation}
Using these results, we obtain the phase shift $\delta\varphi$ for
the photon-statistics-based weak measurement: 
\begin{align}
\delta\varphi & =\int\varphi P_{w}(\varphi)d\varphi-\int\varphi P_{i}(\varphi)d\varphi\nonumber \\
 & =-\lambda\mathrm{Re}[\langle\hat{\sigma}_{z}\rangle_{w}].\label{eq:38}
\end{align}
As seen, the phase shift $\delta\varphi$ is proportional to the real
part of the weak value. From the above results, we obtain: $\mathrm{Re}[\langle\hat{\sigma}_{z}\rangle_{w}]=-\frac{\delta\varphi}{\lambda}$
and $\mathrm{Im}[\langle\hat{\sigma}_{z}\rangle_{w}]=-\frac{\delta n}{2\lambda|\alpha|^{2}\gamma^{2}\left(1-\gamma^{2}\left[\vert\alpha\vert^{4}+2\vert\alpha\vert^{2}+2]\right]\right)}$.
Thus, the real and imaginary parts of the weak value can be extracted
from phase-sensitive displacements $\delta\varphi$ in the optical
field and changes in the average photon number $\delta n$ in our
scheme. This result highlights the universality of weak values in
bridging commutation relations and measurement outcomes across diverse
physical systems. The variance of the photon and phase after measurement 

\begin{subequations}
\begin{align}
(\Delta n)^{2} & =\langle\hat{n}^{2}\rangle_{w}-\langle\hat{n}\rangle_{w}^{2}\\
 & =\kappa^{2}\left(\vert\beta\vert^{6}+6\vert\beta\vert^{4}+7\vert\beta\vert^{2}+1\right)\nonumber \\
 & -\left[\kappa^{2}\left(\vert\beta\vert^{4}+3\vert\beta\vert^{2}+1\right)\right]^{2},\nonumber 
\end{align}
and 

\begin{align}
(\Delta\varphi)^{2} & =\langle\hat{\varphi}^{2}\rangle_{w}-\langle\hat{\varphi}\rangle_{w}^{2}\\
 & =\frac{1}{4}\left[\kappa^{2}(1+\frac{3}{\vert\beta\vert^{2}})\right],\nonumber 
\end{align}

\end{subequations}where $\langle\hat{\varphi}^{2}\rangle_{w}=\int\varphi^{2}P_{w}(\varphi)d\varphi$.
We observe that these variances are associated with $\vert\beta\vert^{2}=\vert\alpha\vert^{2}e^{2\lambda\mathrm{Im}[\langle\hat{\sigma}_{z}\rangle_{w}]}$,
indicating that the variance depends only on the imaginary part of
the weak value. The number-phase uncertainty product $\Delta n\Delta\varphi$
for the SPACS meter after the weak measurement is shown in Fig. \ref{fig:4}.
Figure \ref{fig:4} illustrates that for different interaction strengths
$\lambda$, the number-phase uncertainty product $\Delta n\Delta\varphi$
decreases with increasing $\vert\alpha\vert$ and approaches its minimum
value of 0.5 (see the orange dashed line in Fig. \ref{fig:4}. From
Eq. (\ref{eq:31}), we deduce that $\Delta n\Delta\varphi\geq0.5$,
and this bound is saturated only for the coherent state. When $\vert\alpha\vert$
is large, the state distance between the coherent state and SPACS
becomes negligible. 

In previous study \citep{Bernardo:14}, weak measurement proposal
also had been attempted by taking the photon statistic as a meter
same as our current work. But, in that work they didn't give any specific
details to extract the weak value of measured system observable in
terms of shifts of averages of photon number and phase. To the best
of our knowledge, our current description of weak measurement with
photon statistics which give details analogous to original work of
Aharonov et al. is the first time. If we take $\vert\alpha\vert\gg1$,
the above results can reproduce the results for coherent state case
in Ref. \citep{Bernardo:14}. 

\section{\label{sec:6}Discussion and conclusion}

In this paper, we have provided a comprehensive theoretical analysis
of postselected quantum measurements by considering the photon statistics
of SPACS as a meter. We found that in our proposal, postselected weak
measurements can significantly enhance measurement precision, particularly
in parameter estimation, surpassing the sensitivity achieved with
coherent states. This improvement is quantified through the Fisher
information metric, demonstrating the effectiveness of WVA in weak
interaction regimes. 

Our results show that a comparison with conventional measurement methods
highlights the advantages of postselection, which improves both measurement
sensitivity and accuracy. Additionally, state distance and SNR analyses
reveal that weak measurements can effectively preserve quantum coherence
while boosting precision. We also explicitly introduced a postselected
weak measurement proposal based on a photon-statistics-based meter,
exploring the photon statistics and their associated variances. The
results shows that the real and imaginary parts weak value directly
proportional to phase shift an average photon-number changes after
postselected weak measurement. Moreover, our theoretical calculations
show that for the SPACS meter, the photon number-phase uncertainty
product depends only on the imaginary part of the weak value, revealing
a transformation from SPACS to the coherent-state case as $\vert\alpha\vert$
increases. 

The advantage of a SPACS-based postselected weak measurement over
the coherent-state-based approach is that it enhances measurement
precision, including parameter estimation and SNR, in weak interaction
regimes compared to conventional measurements. As investigated in
Ref. \citep{PhysRevLett.105.010403}, this advantage arises from the
nonclassical nature of SPACS. It is well known that the key to achieving
enhanced precision and SNR in WVA measurements lies in the properties
of the meter. SPACS are more nonclassical than coherent states when
used as a meter \citep{doi:10.1126/science.1103190,PhysRevLett.115.120401,PhysRevLett.128.040503,RN17}
and can be good candidate in associated photon statistics based postselected
measurement precision problems rather than coherent state, even though
it is little \textsl{expensive} quantum resource. Another interesting
point is that our results presented in current work could cover the
related affirmations and claims obtained by coherent state based meter
measurement proposals in Ref. \citep{PhysRevA.106.022619}.

The interaction Hamiltonian used in this work is widely applicable
in quantum optics and circuit QED, as it describes light-atom interactions
in the dispersive regime.l Inspired by the general formalism of weak
measurements introduced by Aharonov et al. \citep{PhysRevLett.60.1351},
we have proposed a weak measurement scheme in which the photon statistics
of light serve as the pointer state. While the original scheme utilized
a canonical position-momentum pair as the pointer variables, our approach
leverages the nontrivial commutation relation between the phase and
photon number to extract both the real and imaginary parts of the
weak value. In our scheme, the real and imaginary parts of the weak
value are extracted by evaluating the shifts in the expectation values
of the photon number and phase, respectively. Our research provide
a novel approach for extracting phase shifts and photon number statistics
of the light field by relating them to the real and imaginary parts
of the weak value of the measured system's observable. Furthermore,
we believe the present theoretical framework, an alternative approach
for improving measurement precision, could apply to precision measurement
and other quantum metrology problems that leverage nonclassical quantum
states.

\subsection*{Funding }

National Natural Science Foundation of China (NSFC) (11865017). 

\subsection*{CRediT authorship contribution statement }

Yi-Fang Ren derived and simulated the theoretical results; Yusuf Turek
wrote the main manuscript text. All authors reviewed the manuscript. 

\subsection*{Declaration of competing interest }

The authors declare that they have no known competing financial interests
or personal relationships that could have appeared to influence the
work reported in this paper. 

\subsection*{Data availability }

No data available. 

\bibliographystyle{apsrev4-1}
\bibliography{Ref}

\end{document}